\documentclass[final,5p,times,twocolumn]{elsarticle}

\usepackage{epsfig}

\usepackage{amssymb}
\usepackage{amsthm}

\usepackage{lineno}

\usepackage{mathtools}
\usepackage{siunitx}
\usepackage{url}
\usepackage{appendix}

\begin{document}
\makeatletter
\def\ps@pprintTitle{%
  \let\@oddhead\@empty
  \let\@evenhead\@empty
  \let\@oddfoot\@empty
  \let\@evenfoot\@oddfoot
}
\makeatother




\title{Quantum Modelling of Magnetism in Strongly Correlated Materials: Evaluating Constrained DFT and the Hubbard Model for Y114}

\author[inst1,inst2,inst3]{Christian Tantardini\corref{cor1}}

\cortext[cor1]{Correspond to christiantantardini@ymail.com}

\affiliation[inst1]{organization={Institute of Physics Belgrade, University of Belgrade, Pregrevica 118, 11080 Belgrade, Serbia.}}

\affiliation[inst2]{organization={Department of Materials Science and NanoEngineering, Rice University, Houston, Texas 77005, United States of America}}

\affiliation[inst3]{organization={Institute of Solid State Chemistry and Mechanochemistry SB RAS, 630128, Novosibirsk, Russian Federation}}

\author[inst3]{Darina Fazylbekova}

\author[inst4]{Sergey V. Levchenko}

\affiliation[inst4]{organization={Skolkovo Institute of Science and Technology, Skolkovo Innovation Center, Bolshoy boulevard 30, Moscow, 121205, Russian Federation}}

\author[inst4,inst5,inst6]{Ivan S. Novikov}

\affiliation[inst5]{organization={Moscow Institute of Physics and Technology, 9 Institutskiy per., Dolgoprudny, Moscow Region, 141701, Russian Federation}}
\affiliation[inst6]{organization={Emanuel Institute of Biochemical Physics RAS, 4 Kosygin Street, Moscow, 119334, Russian Federation}}

\begin{abstract}
Transition-metal compounds represent a fascinating playground for exploring the intricate relationship between structural distortions, electronic properties, and magnetic behaviour, holding significant promise for technological advancements. Among these compounds, YBaCo$_4$O$_{7}$ (Y114) is attractive due to its manifestation of a ferrimagnetic component at low temperature intertwined with distortion effect due to the charge disproportionation on Co ions, exerting profound impact on its magnetic properties. In this perspective paper, we study the structural and magnetic intricacies of the Y114 crystal using a novel first-principles methodology. Traditionally, the investigation of such materials has relied heavily on computational modelling using density-functional theory (DFT) with the on-site Coulomb interaction correction $U$ (DFT+$U$) based on the Hubbard model (sometimes including Hund's exchange coupling parameter $J$, DFT+$U$+$J$) to unravel their complexities.
Herein, we analysed the spurious effects of magnetic-moment delocalisation and spillover to non-magnetic ions in the lattice on electronic structure and magnetic properties of Y114. To overcome this problem we have applied constrained DFT (cDFT) based on the potential self-consistency approach, and comprehensively explore the Y114 crystal's characteristics in its ferrimagnetic order. 
We find that cDFT yields magnetic moments of Co ions much closer to the experimental values than Hubbard model with the parameters $U$ and $J$ fitted to reproduce experimental lattice constants. cDFT allows for an accurate prediction of magnetic properties using oxidation states of magnetic ions as well-defined parameters. 
Through this perspective, we not only enhance our understanding of the magnetic interactions in Y114 crystal, but also pave the way for future investigations into magnetic materials.  
\end{abstract}


\maketitle

\section{Introduction}
\label{sec:Intro}
Co-containing oxides often possess interesting magnetic properties. Among them, YBaCo$_4$O$_{7}$ (abbreviated as Y114) has recently attracted attention \cite{Valldor_2002,VALLDOR2004251,Maignan_2006,Caignaert_2006,Podberezskaya_2013,Chapon_2006,10.1063/1.4792597} due to its complex magnetic behaviour. For example, this structure demonstrates a spin–glass transition at around 66 K from a high-temperature paramagnetic state \cite{Valldor_2002,Maignan_2006,Caignaert_2006,Podberezskaya_2013,Chapon_2006}. 
In such materials, charge disproportionation can occur, which can strongly influence their magnetic properties. Based on the formal oxidation state count, in Y114 three Co ions per primitive unit cell must adopt +2 oxidation state, and one Co ion adopts a +3 oxidation state. Such materials are known to exhibit dynamic distortion due to electronic fluctuations between the Co ions \cite{Valldor_2002,Maignan_2006,Caignaert_2006,Podberezskaya_2013,Chapon_2006}. For this reason, a direct experimental observation of the expected differences in oxygen tetrahedral environments for each unique cobalt site to assign the correct oxidation state is currently not feasible \cite{Valldor_2002,Maignan_2006,Caignaert_2006,Podberezskaya_2013,Chapon_2006}. The inability to refine the structure experimentally due to dynamic distortion can be effectively addressed through a computational approach, as previously demonstrated by Tantardini \textit{et al.} \cite{Tantardini_2018} using Hubbard on-site correction $U$. However, describing magnetism with Hubbard-model-based corrections may not be sufficient due to the lack of experimental data needed to determine the parameters of the model, the Hubbard $U$ and Hund's coupling $J$, which describe the long-range interactions affecting local magnetic moments. 

Here, we investigate the limitations of Hubbard model combined with spin-polarized DFT, incorporating both $U$ and $J$ parameters, using the simplest DFT functional known as the local spin density approximation (LSDA+$U$+$J$) in describing complex magnetic materials, and explain how these materials can be theoretically treated using constrained DFT (cDFT) with potential self-consistency, based on the same LSDA DFT functional. \cite{Gonze_2022}. This approach differs from previously developed cDFT methods \cite{Kaduk_2012,Regan_2016,Wu_2006}, because it imposes charge or magnetic moment hard constraints by finding such a potential that the corresponding self-consistent wavefunctions and electronic density satisfy the constraints exactly, rather than by using a penalty function directly for the deviation of the constrained quantity from the target. Specifically, a Lagrangian potential-based self-consistency constraint \cite{Gonze_2022} will be applied to model cobalt ions in different oxidation states, Co$^{2+}$ and Co$^{3+}$, in Y114. Additionally, the same type of constraint is applied to the magnetic moments of other atoms in the lattice to prevent the spreading of magnetic moments to non-magnetic atoms, such as oxygen. 

In summary, this study adopts an interdisciplinary approach, integrating experimental data obtained by inverse magnetic susceptibility measurements in X-ray powder diffraction at 2 K \cite{Chapon_2006} and advanced computational techniques — specifically leveraging cDFT with a constrained atomic charge and magnetic moments — to unravel the intricate structural, electronic, and magnetic properties of magnetic oxides, emphasising the impact of the dynamic distortions due to the chemical disproportionation on magnetic order.

\begin{figure}
    \centering
    \includegraphics[width=0.3\textwidth]{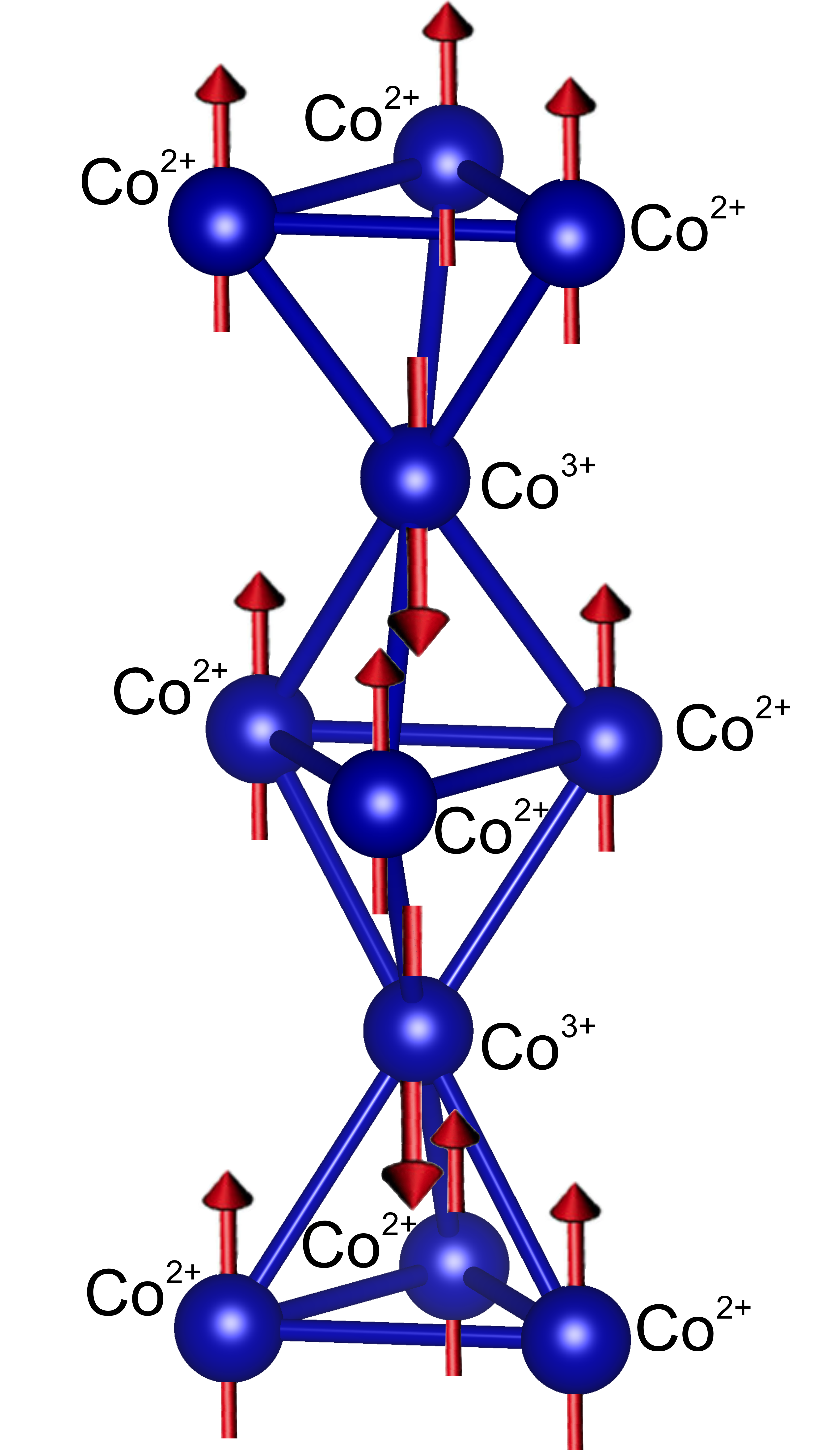}
    \caption{Crystal structure of Y114 with Co$^{2+}$ and Co$^{3+}$ in their ferrimagnetic configuration.}
    \label{fig:Y114_structure}
\end{figure}

\section{Theory}
\label{sec:Theo}
We aim to provide a comparison between Hubbard model and cDFT for describing strongly correlated materials \cite{Himmetoglu2014}. 
In Kohn-Sham (KS) DFT the electronic structure of a material is obtained by solving a system of single-particle equations known as KS equations. The Hamiltonian in these equations includes the sum of kinetic energy and the external potential for a single electron. To these terms, the Hartree term, representing the Coulomb repulsion between all electrons (including the spurious self-interaction), and the exchange-correlation term, approximated in various forms in DFT (e.g., LSDA as considered here), are added.
The KS states $\psi_{k,\nu}$ describe the single electron within a specific band $\nu$ at a specific $k$-point in the reciprocal space. These states are delocalized over the crystal.

In most cases, such a model can fully describe the properties of materials. However, in the case of strongly correlated materials such as transition-metal compounds, the self-interaction error and other exchange-correlation errors in approximate energy-functionals can lead to qualitatively incorrect description of localised valence $d$-orbitals, making it challenging to model oxidation states and magnetic interactions of transition-metal ions in crystals. The same problem occurs in lanthanides and actinides with the localised $f$-orbitals. Over the years, the on-site Coulomb interaction correction $U$ of Hubbard model \cite{anisimov1991band,anisimov1991density,anisimov1993yk} applied to DFT (DFT+$U$) practically addressed this issue by considering specific electronic interactions between atomic orbitals. 
The total energy in DFT+$U$ is formulated as follows:
\begin{align}
    E^{{\rm DFT}+U}[\rho, n] &= E^{{\rm DFT}}[\rho] + E^{U}[n] - E^{dc}[n].
    \label{eq:1}
\end{align}
Here, $\rho$ is the electron density of the system, $n$ is the density matrix for localised atomic orbitals on a specific atom $A$, and $E^{dc}[n]$ is a double-counting term removing the DFT energy contribution of the localized orbitals, which are now described by the Hubbard-like terms. In this framework, two alternative approaches to avoid double counting $(E^{dc})$ have been proposed: the fully localised limit (FLL) \cite{Liechtenstein1995} and the around mean-field (AMF) \cite{Czyzyk1994}. The density matrix is generated by projecting KS states onto atomic orbitals with specific angular momentum $\ell$ and associated momentum projection $m$ of the atom $A$:

\begin{align}
    n^{A\sigma}_{mm'} &= \sum_{k,\nu} f_{k \nu}^{\sigma}  \langle \varphi^{A\sigma}_{m} | \psi_{k,\nu} \rangle \langle \psi_{k,\nu} | \varphi^{A\sigma}_{m'} \rangle,
\end{align}

\noindent where $f$ is the Fermi-Dirac distribution, $\sigma$ is the spin on the atom $A$, and $\varphi$ are the atomic orbitals described as product of radial functions and spherical harmonics centered on the atoms.
If only diagonal terms of the local density matrix $n^{A\sigma}_{mm'}$ are considered, the term $E^{U}[n]$ will lose its invariance under rotation. Therefore, the terms that come from the off-diagonal local density matrix should be also considered.

The energy term in the Hubbard model ($E^{U}[n]$) is given by:

\begin{align}
    E^{U}[n] &= \frac{1}{2} \sum_{A=1}^{M} \sum_{\{m\},\sigma} \bigg\{ \bigg\langle \varphi^{A}_{m},\varphi^{A}_{m''} \bigg| V_{ee} \bigg| \varphi^{A}_{m'}, \varphi^{A}_{m'''} \bigg\rangle n^{\sigma}_{mm'} n^{-\sigma}_{m''m'''} \nonumber \\
    & + \bigg( \bigg\langle \varphi^{A}_{m},\varphi^{A}_{m''} \bigg| V_{ee} \bigg| \varphi^{A}_{m'}, \varphi^{A}_{m'''} \bigg\rangle \nonumber \\
    & - \bigg( \bigg\langle \varphi^{A}_{m},\varphi^{A}_{m''} \bigg| V_{ee} \bigg| \varphi^{A}_{m'''}, \varphi^{A}_{m'} \bigg\rangle \bigg) n^{\sigma}_{mm'} n^{\sigma}_{m''m'''} \bigg\}.
\end{align}
Here, $V_{ee}$ is the screened electron-electron Coulomb repulsion. In the FLL formulation of $E^{dc}$ seen in the Eq.~\ref{eq:1}, the parameters $U$ and $J$, referred to as screened Coulomb and Hund's coupling parameters, enter as follows:

\begin{align}
    E^{dc}[n] = \frac{1}{2}U N (N-1) - \frac{1}{2} J[N^{\uparrow}(N^{\uparrow} - 1) + N^{\downarrow}(N^{\downarrow} - 1)]
\end{align}
where $N^\sigma=\sum_m n^{A\sigma}_{mm}$, and $N = \sum_\sigma N^\sigma$.
In this formulation, which slightly differs from AMF, a constraint is applied to specific atomic orbitals of an atom. In practice, an energy constraint is applied to the Coulomb interaction between specific atomic orbitals of a chosen atom, preventing them from spreading over the entire structure.
The DFT+$U$+$J$ method shares similarities with the Hartree-Fock (HF) method. It essentially replaces certain electronic interactions with a Hamiltonian reminiscent of HF Hamiltonian, similar to hybrid functionals, where part of the functional involves a Fock exchange operator acting on KS states. However, DFT+$U$+$J$ differs by utilising screened effective interactions and focusing only on a specific subset of states.

Within DFT+$U$+$J$ an assumption of orbital independence is made due to the localised nature of the orbitals the correction is applied to. Despite its formal resemblance to HF, DFT+$U$+$J$ operates on KS wave functions, lacking a direct physical interpretation beyond reproducing the charge density. In essence, DFT+$U$+$J$ bridges concepts from HF and hybrid functionals, incorporating screened interactions and orbital decoupling while selectively applying corrections to specific states in the system.

Moreover, $U$ and $J$ can be construed as parameters of the Hubbard model representing the weight of an additional penalty function integrated into the total energy. This augmentation introduces a biased solution to DFT. The values of these parameters are intricately linked to the atomic environment and concentration of specific atoms relative to the overall quantity of atoms within the given structure. Their determination necessitates the application of one of four distinct methodologies: (i) fitting, wherein various properties such as lattice parameters or magnetic moments are juxtaposed for different $U$ and $J$ values at varying concentrations of strongly correlated atom types, (ii) the linear response approach, colloquially known as the Cococcioni-Gironcoli method \cite{cococcioni2005linear,TIMROV2022108455,PhysRevB.103.045141}, (iii) the constrained random phase approximation (cRPA) \cite{Amadon2014}, or (iv) pseudohybrid Hubbard density functional ACBN0 \cite{PhysRevX.5.011006}.
Regrettably, (ii)-(iv) are notably intricate and do not always yield results close to experiment, and (i) cannot be employed \textit{a priori} without experimental values for comparative analysis. Furthermore, as we demonstrate below, LSDA+$U$+$J$ fails to describe the charge and magnetic moment distribution correctly in some cases. 

The $U$ and $J$ parameters play pivotal role in determining the magnetic moments ($\mu$) of materials, particularly in systems characterised by strong electron-electron correlation and localised electronic states. These parameters influence magnetic moments through their effects on electronic configurations and spin alignments within the material's electronic structure.

\begin{enumerate}
    \item \textbf{Hubbard $U$ Parameter}:
    \begin{itemize}
        \item The Hubbard $U$ parameter characterises the on-site Coulomb repulsion between electrons occupying the same atomic orbital. It is quantified by the Hubbard Hamiltonian term:
        \[ \hat{H}_U = U \sum_{i} \hat{n}_{i\uparrow} \hat{n}_{i\downarrow} \]
        where $\hat{n}_{i\sigma}$ represents the number operator for electrons with spin $\sigma$ on site $i$.
        \item Increasing $U$ leads to a stronger repulsion between electrons on the same orbital, promoting electron localisation on different orbitals. This effect is captured by the Hubbard Hamiltonian.
        \item In magnetic materials, localised electrons tend to align their spins to minimise the Coulomb repulsion energy, contributing to the material's magnetic moment. Therefore, larger values of $U$ generally result in stronger electron localisation and larger magnetic moments.
    \end{itemize}
    
    \item \textbf{Hund's Coupling Parameter $J$}:
    \begin{itemize}
        \item Hund's coupling $J$ represents the exchange interaction between electrons with parallel spins on the same atomic site. It favours parallel spin configurations over anti parallel ones and is described by the Hamiltonian term:
        \[ \hat{H}_J = -J \sum_{i} \left( \hat{S}_{i}^{2} - \frac{\hat{n}_{i}(\hat{n}_{i} - 1)}{2} \right) \]
        where $\hat{S}_i$ is the total spin operator and $\hat{n}_i$ is the total electron number operator on site $i$.
        \item Increasing $J$ enhances the energy benefit of aligning spins, particularly in high-spin configurations with multiple unpaired electrons occupying orbitals with the same angular momentum ($d$- or $f$-orbitals), but different momentum projections $m$. This effect stabilises high-spin states and contributes to larger magnetic moments in magnetic materials.
    \end{itemize}
    
    \item \textbf{Interaction between $U$ and $J$}:
    \begin{itemize}
        \item $U$ and $J$ parameters often exhibit synergistic effects, where a larger $U$ can enhance the effectiveness of $J$ in stabilising high-spin states.
        \item However, there can also be competing effects between $U$ and $J$. For example, while larger values of $U$ generally lead to more localised electron states and larger magnetic moments, excessively large $U$ can hinder electron mobility and suppress magnetic ordering.
    \end{itemize}
    
    \item \textbf{Material-Specific Considerations}:
    \begin{itemize}
        \item The influence of $U$ and $J$ parameters on magnetic moments depends on the material's specific electronic structure, crystal symmetry, and other material parameters.
        \item Determination of $U$ and $J$ parameters tailored to the material of interest is essential for accurate predictions of magnetic properties. This can be achieved through empirical fitting or theoretical calculations based on electronic-structure methods, as discussed above.
    \end{itemize}
\end{enumerate}

The above mentioned drawbacks can be overcome through the advanced potential-based self-consistency constrained Density Functional Theory (cDFT) energy functional \cite{Gonze2022} denoted here as \(E^{\rm cDFT}\). This functional, designed to admit the same self-consistent solution as given by specific equations, is expressed as follows:

\begin{equation} \label{EcDFT}
E^{\rm cDFT}_{v_{\rm ext},N_{\rm A}}[u] = E^{\rm v}_{v_{\rm ext}}[u] - R^{\rm v}_{\rm A}[u] (W_{\rm AA})^{-1} \left(\rho^{\rm v}_{\rm A}[u] - N_{\rm A}\right)
\end{equation}
Here, $E^{\rm v}_{v_{\rm ext}}[u]$ is the  DFT energy, $R^{\rm v}_{\rm A}[u]$ is the residual self-consistent potential, $W_{\rm AA}$ is the integral of a weight function $w_{\rm A}(\mathbf{r})$ (which is 1 inside the volume associated with fragment A, and 0 outside) squared, \(u\) represents the screened potential, \(v_{\rm ext}\) the external potential depending on atomic positions and cell parameters, and \(N_{\rm A}\) is the number of electrons associated with a specific atomic fragment \(A\). Notably, both \(v_{\rm ext}\) and \(N_{\rm A}\) are treated as external parameters in the calculation. The residual self-consistent potential is the integral of residual potential (i.e., the difference between the output and input screening potentials) times the weight function.

\begin{table}[h]
    \centering
    \begin{tabular}{c|c|c|c|c|c}
     Atom      &       x      &     y       &     z       &   LSDA+$U$+$J$ /  &     cDFT /        \\ 
               &              &             &             &      $\mu_B$     &    $\mu_B$         \\
     Ba        &     0.3333   &   0.6667    &   0.9816    &    -0.0003       &    0.0000          \\
     Ba        &     0.6667   &   0.3333    &   0.4816    &    -0.0003       &    0.0000          \\
     Y         &     0.3333   &   0.6667    &   0.3780    &     0.0037       &    0.0000          \\
     Y         &     0.6667   &   0.3333    &   0.8780    &     0.0037       &    0.0000          \\
     Co2       &     0.1692   &   0.3385    &   0.6908    &     2.4181       &    1.9974          \\
     Co2       &     0.6615   &   0.8308    &   0.6908    &     2.4181       &    1.9974          \\
     Co2       &     0.1692   &   0.8308    &   0.6908    &     2.4181       &    1.9974          \\
     Co2       &     0.8308   &   0.6615    &   0.1908    &     2.4181       &    1.9974          \\
     Co2       &     0.3385   &   0.1692    &   0.1908    &     2.4181       &    1.9974          \\
     Co2       &     0.8308   &   0.1692    &   0.1908    &     2.4181       &    1.9974          \\
     Co1       &     0.0000   &   0.0000    &   0.4423    &    -2.5309       &   -3.3055          \\
     Co1       &     0.0000   &   0.0000    &   0.9423    &    -2.5309       &   -3.3055          \\
     O         &     0.0091   &   0.5045    &   0.7642    &     0.0269       &    0.0000          \\
     O         &     0.4955   &   0.5045    &   0.7642    &     0.0269       &    0.0000          \\
     O         &     0.4955   &   0.9909    &   0.7642    &     0.0269       &    0.0000          \\
     O         &     0.9909   &   0.4955    &   0.2642    &     0.0269       &    0.0000          \\
     O         &     0.5045   &   0.4955    &   0.2642    &     0.0269       &    0.0000          \\
     O         &     0.5045   &   0.0091    &   0.2642    &     0.0269       &    0.0000          \\
     O         &     0.1634   &   0.3267    &   0.5033    &    -0.1161       &    0.0000          \\
     O         &     0.6733   &   0.8366    &   0.5033    &    -0.1161       &    0.0000          \\
     O         &     0.1634   &   0.8366    &   0.5033    &    -0.1161       &    0.0000          \\
     O         &     0.8366   &   0.6733    &   0.0033    &    -0.1161       &    0.0000          \\
     O         &     0.3267   &   0.1634    &   0.0033    &    -0.1161       &    0.0000          \\
     O         &     0.8366   &   0.1634    &   0.0033    &    -0.1161       &    0.0000          \\
     O         &     0.0000   &   0.0000    &   0.2571    &    -0.0658       &    0.0000          \\
     O         &     0.0000   &   0.0000    &   0.7571    &    -0.0658       &    0.0000          \\ 
    \end{tabular}
    \caption{Atomic fractional coordinates for the ferrimagnetic Y114 structure optimised with LSDA+$U$+$J$ (lattice parameters: $a = b = 6.274715382$ \AA; $c = 10.234961467$ \AA; $\alpha = \beta = 90$; $\gamma = 120$). Magnetic moments calculated with LSDA+$U$+$J$ and constrained DFT (cDFT). The relaxed atomic positions, primitive vectors, and magnetic moments are all within 10$^{-4}$ of the expected values based on crystal symmetry.}
    \label{tab:my_label}
\end{table}

The functional \(E^{\rm cDFT}_{v_{\rm ext},N_{\rm A}}[u]\) is stationary at the self-consistent potential \(v^*\), leading to the following self-consistency relation:

\begin{equation} \label{ESCcDFT}
E^{\rm SC}_{v_{\rm ext},N_{\rm A}} = E^{\rm cDFT}_{v_{\rm ext},N_{\rm A}}[v^*]
\end{equation}

Moreover, the functional stationary behaviour is expressed as:

\begin{equation} \label{EcDFT_stationary}
E^{\rm cDFT}_{v_{\rm ext},N_{\rm A}}[u] = E^{\rm cDFT}_{v_{\rm ext},N_{\rm A}}[v^*] + \mathcal{O}\big((u-v^*)^2 \big)
\end{equation}

The gradient of this functional with respect to the screened potential \(u\) is given by:

\begin{align} \label{dEcdft_u}
\frac{\delta E^{\rm cDFT}_{v_{\rm ext},N_{\rm A}}[u]}{\delta u(\mathbf{r})} & = \int \chi_0(\mathbf{r},\mathbf{r}') R^{+{\rm v}}[u,\Lambda_{\rm A}[u]](\mathbf{r}') d\mathbf{r}' \nonumber \\
& + \left( \int \epsilon_e(\mathbf{r},\mathbf{r}') w_{\rm A}(\mathbf{r}') d\mathbf{r}' \right) (W_{\rm AA})^{-1} \big(\rho^{\rm v}_{\rm A}[u] - N_{\rm A}\big)
\end{align}
Here, \(\Lambda_{\rm A}[u]\) is defined as:

\begin{equation} \label{Lambda_A}
\Lambda_{\rm A}[u] \triangleq -R^{\rm v}_{\rm A}[u] (W_{\rm AA})^{-1},
\end{equation}
$\epsilon_e(\mathbf{r},\mathbf{r}')$ is the electron dielectric response function, and $\chi_0(\mathbf{r},\mathbf{r}')$ is the independent-particle susceptibility. The formulation introduces a residual for cDFT, denoted as \(R^{\rm cDFT}\), defined as:

\begin{align} \label{RcDFT}
R^{\rm cDFT}[u](\mathbf{r}') & = R^{\rm v}[u]({\bf r}') + \Lambda w_{\rm A}({\bf r}') \nonumber \\ 
& + c \, w_{\rm A}(\mathbf{r}') \left(\rho_{\rm A}[u] - N_{\rm A}\right)
\end{align}
where $c$ is a constant, whose value is formally arbitrary, but for practical purposes 
should be of order one, as it defines the balance between the convergence inside the space spanned by $w_{\rm A}$ and the convergence inside the space perpendicular to it. 
The cDFT residual vanishes when both \(R^{\rm v}\) and \(\rho^{\rm v}_{\rm A}[u] - N_{\rm A}\) vanish, indicating self-consistency.

The formulation allows for easy computation of derivatives and forces within the cDFT framework, using the \(2n+1\) theorem of perturbation theory. Notably, derivatives with respect to number of electrons (\(N_{\rm A}\)) yield quantities such as the chemical potential of fragment \(A\) (\(\chi_A\)).

The advantages of cDFT for describing magnetic moments are the following:

\begin{enumerate}
    \item \textbf{Targeted Studies:} cDFT allows one to perform targeted studies by imposing specific constraints on the electronic density, such as fixing the magnetic moments of certain atoms or regions within a material. This capability is particularly useful for investigating phenomena like magnetic phase transitions, spin-crossover materials, or the effects of magnetic doping on electronic properties.
    
    \item \textbf{Understanding Magnetism at the Atomic Level:} With cDFT, it is possible to explore the atomic-scale origins of magnetism in materials. By controlling the magnetic moments of individual atoms or groups of atoms, one can dissect the contributions of different electronic orbitals and chemical environments to the overall magnetic behaviour. This level of detail is crucial for understanding the microscopic mechanisms driving magnetic phenomena.
    
    \item \textbf{Consistent Description of Magnetic Interactions Between Localized Magnetic Moments:} Magnetic interactions play a fundamental role in determining the properties of magnetic materials. The advantage of the hard constraint in cDFT is the full self-consistency of the electronic states fulfilling the constraint. Thus, interaction between the magnetic moments localised according to the constraint is described consistently with interactions that cause the localisation.

\begin{figure}
    \centering
    \includegraphics[width=0.51\textwidth]{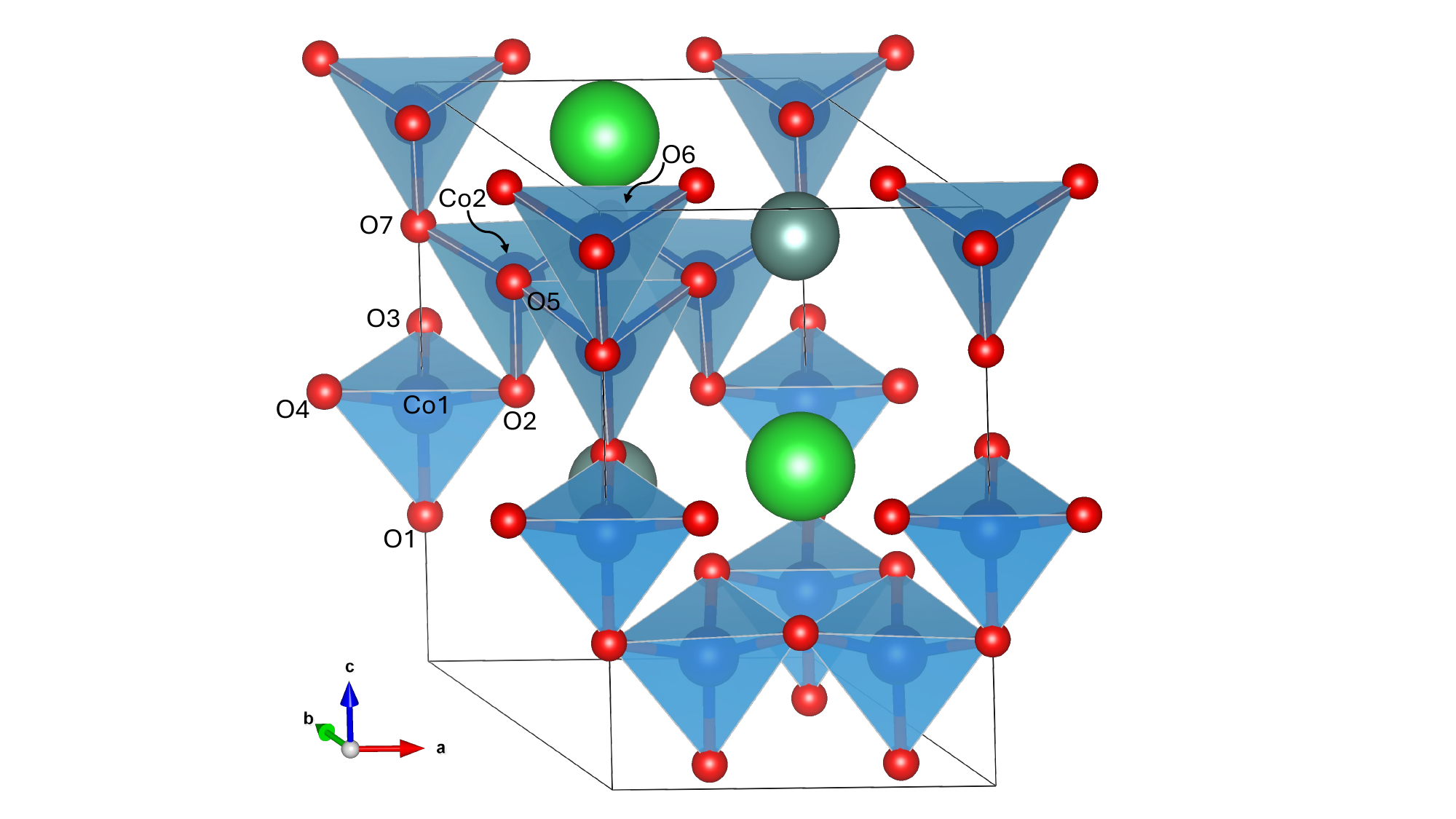}
    \caption{Crystal Structure of the optimised ferrimagnetic Y114. Legend: Co within the tetrahedron, azure; Ba, green; Y, grey; O, red.}
    \label{fig:crys-ferri}
\end{figure}
    
    \item \textbf{Prediction of Magnetic Properties:} By solving the electronic structure problem self-consistently within the constraints imposed by cDFT, one can obtain accurate predictions of various magnetic properties, such as magnetic moments, magnetic susceptibilities, and magnetic exchange interactions. These predictions can be compared with experimental measurements, providing valuable insights into the underlying physics of magnetism in materials.
    
    \item \textbf{Exploration of Complex Magnetic Systems:} cDFT can be applied to explore the magnetic properties of complex systems, including magnetic nanoparticles, magnetic thin films, and magnetic heterostructures. These systems often exhibit intricate magnetic behaviours arising from size effects, interface effects, or proximity-induced magnetism. cDFT enables one to unravel these complexities and understand how they influence the overall magnetic behaviour of the system.
    
    \item \textbf{Design of New Magnetic Materials:} By leveraging the predictive capabilities of cDFT, one can accelerate the discovery and design of new magnetic materials with tailored magnetic properties. By systematically exploring the parameter space of different magnetic configurations, compositions, and structures, cDFT-guided computational screening can identify promising candidates for experimental synthesis and characterisation.
\end{enumerate}

As a test example for cDFT theoretical framework, we investigate atomic and electronic structure, and magnetic interactions in Y114. 
Our choice of Y114 as a test case is strategic for several reasons. Firstly, its complex crystal structure and rich electronic properties make it an ideal candidate for exploring the interplay between electronic correlations, lattice distortions, and magnetic interactions. Additionally, the compound exhibits metallic behaviour with the drop of conductivity with increasing of temperature \cite{Tsipis_2005} and antiferromagnetic component  at low temperature \cite{Valldor_2002,Huq_2006,Chapon_2006}. 
 By utilising DFT+$U$+$J$ and cDFT methods, we aim to capture the subtle interplay between electron-electron interactions and structural distortions that underlie the magnetism in Y114. These methods allow us to incorporate the effects of strong electron correlation and spin-orbit coupling, providing a more accurate description of the material's electronic structure compared to conventional DFT approximations.

Through comparative analysis of our theoretical predictions with experimental observations, we validate and refine our theoretical framework. In the future we expect to extend our investigation to other magnetic materials.

In essence, our study of the magnetism in Y114 serves is a stepping stone towards a more comprehensive theoretical understanding of complex materials, with implications for diverse fields ranging from condensed matter physics to materials science and beyond.

\section{Computational Details}
\label{sec:CompDet}
We have performed spin-polarised collinear magnetic calculations along z-axis considering opposite initial spin magnetic moments orientation for Co$^{2+}$ respect to Co$^{3+}$.
The atomic positions and lattice parameters of the hexagonal crystal structure of Y114 (Materials Project number: mp-19151) are fully relaxed without spin-orbit coupling. This optimisation was executed in the plane-waves basis (PW) framework, employing the Hubbard model applied to LSDA DFT functional  (LSDA+$U$+$J$) \cite{Perdew1992a,anisimov1991band,anisimov1991density,anisimov1993yk}. 
We have fitted the $U$ and $J$ parameters by comparing the computed lattice constants with those from neutron diffraction experiment on Y114 \cite{Valldor_2002}. The final deviations from experimental values of lattice constants $a$ and $c$ were -0.27\% and 0.34\%, respectively, obtained with $U = 8$ eV for Co$^{2+}$, $U = 6$ eV for Co$^{3+}$, and $J = 0.1$ eV for both Co$^{2+}$ and Co$^{3+}$ with slightly different tetrahedral coordination.
Potential-based self-consistency cDFT \cite{Gonze_2022} with the LSDA functional without spin-orbit coupling was used to self-consistently optimise the magnetic moments for the ferrimagnetic structure, previously optimised with Hubbard model (see Fig.~\ref{fig:Y114_structure}), constraining the charge of cobalt atoms to +3 and +2 for corresponding coordinations, and magnetic moments of Y, Ba, and O to zero. The radii of 2 Bohr (the span of the weight functions $w_{\rm A}(\mathbf{r})$) were chosen to calculate the spherical integrals around atoms. The weight function goes smoothly from 1 to 0 in the region from  1.8 to 2.0 Bohr.
This type of partitioning is specific for optimised norm-conserving Vanderbilt pseudopotentials (ONCVPs), which were taken from PseudoDojo project \cite{VANSETTEN201839,PhysRevB.88.085117} \url{pseudo-dojo.org}.
The PW kinetic energy cut-off was set at 50 Ha, with a $6 \times 6 \times 3$ $\Gamma$-centered $k$-point grid. Convergence was reached when forces fell below $5 \cdot 10^{-5}$ Ha/Bohr.
For the geometry optimisation with LSDA+$U$+$J$ we used Quantum Espresso v.7.2 \cite{Giannozzi_2009,Giannozzi_2017}, while cDFT calculations were executed utilising the \texttt{ABINIT} code \cite{Gonze2016,Gonze_2020,Romero2020}.

\begin{figure*}
    \centering
    \includegraphics[width=1.0\textwidth]{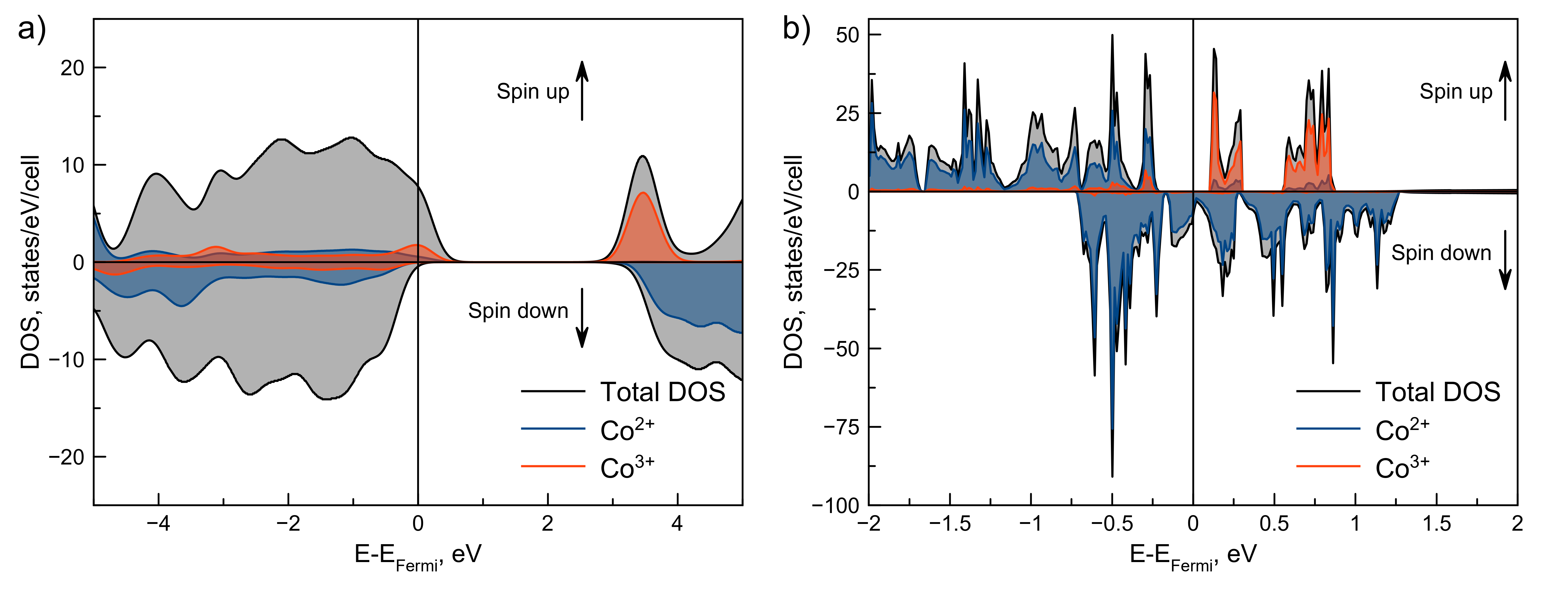}
    \caption{Total and Co 3$d$-projected density of states (pDOS) calculated with LSDA+$U$+$J$ (a) and cDFT (b).}
    \label{fig:pDOS}
\end{figure*}

\section{Results and Discussion}
\label{sec:ResDis}
We fully relaxed the hexagonal structure of Y114 with the ferrimagnetic order, adjusting both atomic positions and lattice constants, characterised by space group $P6_{3}mc$. Several optimisations were conducted using the LSDA+$U$+$J$ approach, applying different values of $U$ and $J$ on a grid to the 3$d$-electrons of Co ions. This iterative process aimed to determine the optimal $U$ and $J$ values that align the computed lattice constants with the experimental values obtained via neutron diffraction at 10 K, as reported in the literature \cite{Valldor_2002}, see the Computational Details section.

The Y114 crystal with the ferrimagnetic order exhibits two slightly different environments for Co$^{2+}$ to Co$^{3+}$ as seen in Fig.~\ref{fig:crys-ferri}. This was not possible to observe in some experiments due to the electron delocalisation between cobalt sites, inducing fluctuations in the Co oxidation states from Co$^{2+}$ to Co$^{3+}$ and vice versa\cite{Valldor_2002,Chapon_2006,Soda_2006,Huq_2006}. 

\begin{table}[ht]
    \centering
    \begin{tabular}{c|c|c|c}
     Bond         &       exp. \cite{Valldor_2002}    &   GGA+$U$ \cite{Tantardini_2018}      &   LSDA+$U$+$J$ \\
                  &              &  (no magnetic)  &  (ferrimagnetic) \\
    \hline     
     Co1-O1       &    2.0339    &   1.9344   &   1.8955      \\
     Co1-O2       &    1.9239    &   1.8983   &   1.8824      \\
     Co1-O3       &    1.9239    &   1.8983   &   1.8818      \\
     Co1-O4       &    1.9239    &   1.8983   &   1.8818      \\
     Co2-O2       &    1.8875    &   1.8752   &   1.92581     \\
     Co2-O5       &    1.9057    &   1.9325   &   1.92581     \\
     Co2-O6       &    1.9057    &   1.9325   &   1.92581     \\
     Co2-O7       &    1.9368    &      -     &   1.96010     \\
    \end{tabular}
    \caption{Bond lengths in \AA ~for Y114 structure shown in the Fig.~\ref{fig:crys-ferri} obtained from experiment \cite{Valldor_2002}, GGA+$U$ non-magnetic state calculations  \cite{Tantardini_2018}, and ferrimagnetic LSDA+$U$+$J$ calculations performed in this work.}
    \label{tab:bond}
\end{table}

The coordination of Co$^{3+}$ (see Fig.~\ref{fig:crys-ferri}) is characterised by a distorted tetrahedral geometry in the experimental structure \cite{Valldor_2002}.
In this structure, the apical bond (Co1-O1) is the longest one between Co and O (see Table~\ref{tab:bond}), while the bonds with the other three basal oxygen atoms have nearly identical lengths (see Co1-O2, Co1-O3, and Co1-O4 in Table~\ref{tab:bond}).
In both the non-magnetic structure studied by Tantardini \textit{et al.} \cite{Tantardini_2018} and the ferrimagnetic structure investigated here, a distorted tetrahedron associated with Co$^{3+}$ is observed (see Table~\ref{tab:angle}). In these computed structures, the longest bond remains the apical one. 
However, in the ferrimagnetic structure, the two basal oxygen atoms that form bridges between Co$^{3+}$ and Ba (see Co1-O3 and Co1-O4 in Table~\ref{tab:bond}) have similar bond length, and they slightly differ from the one for the third basal oxygen atom, which forms a bridge between Co$^{3+}$ and Y (see Co1-O2 in Table~\ref{tab:bond}).

\begin{table}[ht]
    \centering
    \begin{tabular}{c|c|c|c}
     Bond         &       exp. \cite{Valldor_2002}    &   GGA+$U$ \cite{Tantardini_2018}               &   LSDA+$U$+$J$ \\
     Angle        &              &  (no magnetic)  &  (ferrimagnetic) \\
    \hline     
    O1-Co1-O2     &  108.5881   &  117.6609   &    109.3700   \\  
    O1-Co1-O3     &  108.5881   &  100.1790   &    109.3755   \\  
    O1-Co1-O4     &  108.5881   &  100.1791   &    109.3755   \\  
    O2-Co1-O3     &  110.3398   &  117.6609   &    109.5612   \\      
    O2-Co1-O4     &  110.3398   &  117.6609   &    109.5612   \\      
    O3-Co1-O4     &  110.3398   &  100.1790   &    109.5833   \\      
    O2-Co2-O5     &  112.9615   &     -       &    113.9086   \\  
    O2-Co2-O6     &  112.9615   &     -       &    113.9086   \\  
    O2-Co2-O7     &  107.0432   &     -       &    108.3735   \\  
    O5-Co2-O6     &  103.2645   &     -       &    104.8207   \\ 
    O5-Co2-O7     &  110.3193   &     -       &    107.7532   \\ 
    O6-Co2-O7     &  110.3193   &     -       &    107.7533   \\
    \end{tabular}
    \caption{Bond angles in degrees for experimental Y114 structure shown in the Fig.~\ref{fig:crys-ferri} \cite{Valldor_2002}, GGA+$U$ non-magnetic structure from the previous work by Tantardini \textit{et al.} \cite{Tantardini_2018}, and the ferrimagnetic LSDA+$U$+$J$ structure calculated in this work.}
    \label{tab:angle}
\end{table}

The O-coordination of Co$^{2+}$ (see Fig.~\ref{fig:crys-ferri}) is a distorted tetrahedron in both the experimental and ferrimagnetic structures, but it is triangular in the non-magnetic structure, as shown by Tantardini \textit{et al.} \cite{Tantardini_2018} (see Table~\ref{tab:bond} and Table~\ref{tab:angle}).
In the experimental structure, the apical distance between Co$^{2+}$ and the oxygen that bridges Y and Co$^{3+}$ is the shortest (see Co2-O5 in Table~\ref{tab:bond}). However, in the ferrimagnetic structure obtained here using LSDA+$U$+$J$, this distance is identical to that between Co$^{2+}$ and the oxygen atoms bridging the other Co$^{2+}$ ions in the Kagome lattice (see Co2-O6 and Co2-O7 in Table~\ref{tab:bond}).
Furthermore, in both the experimental and calculated ferrimagnetic structures, the longest bond is between Co$^{2+}$ and the oxygen that bridges to Co$^{3+}$.
This distance is increased to 2.19 \AA~by the triangular distortion of Co$^{2+}$ in the non-magnetic structure obtained by Tantardini \textit{et al.} \cite{Tantardini_2018}.

In summary, our study successfully describes the different atomic sites within the material.
Cobalt atoms are arranged such that Co$^{3+}$ sites are positioned between the triangles formed by Co$^{2+}$ sites, characteristic of the Kagome lattice structure \cite{Ghimire2020,10.1143/ptp/6.3.306,kiesel2012sublattice}.

The ferrimagnetic properties of the structure (see Table~\ref{tab:my_label}) are due to distinct magnetic moments for the two types of Co sites. The cobalt (Co) atom has an atomic number of 27 and a valence electronic configuration of $4s^{2}3d^{7}$. In a tetrahedral coordination, the Co ion has three $3d$ orbitals that do not participate in chemical bonding. These orbitals are divided into two different groups based on symmetry: the $t_{2}$ group, consisting of the $d_{xy}$, $d_{yz}$, and $d_{xz}$ orbitals, which are higher in energy compared to the $e$ group, consisting of the $d_{x^{2}-y^{2}}$ and $d_{z^{2}}$ orbitals \cite{jean2005molecular}.
Thus, for Co$^{2+}$, the valence electronic configuration changes to $4s^{0}3d^{7}$, which can only exist in one possible configuration with 3 unpaired spin magnetic moments (i.e., 3 $\mu_{B}$).
In contrast, for Co$^{3+}$, the valence electronic configuration changes to $4s^{0}3d^{6}$, which can adopt two configurations: a high-spin state with 4 unpaired electrons (i.e., 4 $\mu_{B}$) or a low-spin state with 2 unpaired electrons (i.e., 2 $\mu_{B}$).
The LSDA+$U$+$J$ magnetic moments are 2.41 $\mu_{B}$ for Co$^{2+}$ and -2.53 $\mu_{B}$ for Co$^{3+}$. The deviation between LSDA+$U$+$J$ magnetic moments and the expected values can be explained by a spillover of magnetic moment to other atoms in the crystal (e.g., oxygen), by Co ions adopting lower-spin states, by interaction between Co ions, or by a combination of these factors. We find a rather small magnetisation of the order of 0.1 $\mu_{B}$ on the oxygen atoms, which cannot explain the deviation. The experimental values of magnetic moments, deduced from inverse magnetic susceptibility measurement in X-ray powder diffraction at 2 K \cite{Chapon_2006}, are -3.49(8) $\mu_{B}$ for Co$^{3+}$ and 2.19(4) $\mu_{B}$ for Co$^{2+}$. Interestingly, the experimental magnetic moment of Co$^{2+}$ is even further from the expected high-spin moment 3 $\mu_{B}$ than the LSDA+$U$+$J$, and falls between the high-spin and low-spin value (1 $\mu_{B}$) for the free Co$^{2+}$ ion. As discussed in the literature on quantum chemical topology of spin-density distributions \cite{bruno2020spin,macetti2018spin}, this can be explained by the interaction between Co ions in the lattice.

To address the discrepancies in magnetic moments derived from LSDA+$U$+$J$ and experiments, we employed cDFT \cite{Gonze_2022} imposing a constraint on charge of the cobalt atoms. Based on the previous works \cite{Tantardini_2018,Valldor_2002,Chapon_2006,Soda_2006,Huq_2006}, three of the four Co atoms per formula unit were assigned the oxidation state of +2, and the remaining Co atom was constrained to the oxidation state of +3. Furthermore, oxygen barium, and yttrium are expected to be non-magnetic, and therefore their magnetic moments were constrained to be zero.
It is noteworthy that cDFT is parameterised directly to account for local effects attributable to the different oxidation states of cobalt atoms, while the $U$ and $J$ parameters in traditional LSDA+$U$+$J$ are chosen based on other criteria, and are forced to be the same for magnetic ions in different oxidation states. This results in delocalization of the $d$-electrons and incorrect magnetic moments. cDFT yields the magnetic moments of -3.30 $\mu_{B}$ for Co$^{3+}$ and 2.00 $\mu_{B}$ for Co$^{2+}$, which are much closer to the experimental values \cite{Chapon_2006} than LSDA+$U$+$J$ (see  Table~\ref{tab:my_label}). Thus, cDFT correctly reproduces the interaction (via chemical bonding) between Co ions, resulting in the apparent reduction of local spin moment on Co ions, particularly on Co$^{2+}$ with $(d^{7})$ configuration.

These differences between LSDA+$U$+$J$ and cDFT are further investigated by examining the projected density of states (pDOS), as illustrated in Figure ~\ref{fig:pDOS}. We find drastic differences between the two methods. In contrast to LSDA+$U$+$J$, which exhibits spin-majority states with predominant O 2$p$ character at and near the Fermi energy (as depicted in Fig.~\ref{fig:pDOS}a), cDFT pDOS in general exhibits sharper peaks, indicating state localisation, and the states around the Fermi level are spin-minority states with predominant Co$^{2+}$ 3$d$ character (Fig.~\ref{fig:pDOS}b). The smaller contribution of O 2$p$ states around the Fermi level in the case of cDFT can be attributed to the constraint steering the magnetic moments on non-magnetic ions in the crystal to zero. Thus, cDFT provides a unique and, according to the experimental results, more accurate representation of electronic structure and orbital occupation within Y114.

\section{Conclusions}
\label{sec:Conc}
In this study, we have demonstrated how potential-based self-consistent cDFT can be used to improve description of magnetic interactions in complex magnetic compounds, using Y114 as a prototypical example. While Hubbard model, with the parameters $U$ and $J$ fitted to reproduce experimental lattice constants, correctly predicts charge disproportionation leading to slightly different tetrahedral O-coordination of Co$^{2+}$ and Co$^{3+}$ ions, and ferrimagnetic order, it fails to correctly describe magnetic moments of the Co ions. By imposing potential-based self-consistent charge constraints on the Co ions, and constraining the magnetic moments of non-magnetic ions (Y, Ba, and O) to be zero in the cDFT framework, we obtain magnetic moments of Co ions much closer to experimental values. Thus, potential-based self-consistent cDFT allows for accurate prediction of magnetic properties using the much more intuitive parameter choice (charges around the magnetic ions) than choice of $U$ and $J$ in Hubbard model. 

The cDFT results confirm the value of magnetic moment of Co$^{2+}$ ions close to 2 $\mu_{B}$, which is exactly between high-spin and low-spin states of an isolated Co$^{2+}$ ion. This is explained by a strong interaction (bonding) between Co ions in the lattice. This bonding can also explain the dynamic redistribution of the +2/+3 charge in the Co lattice and the resulting oxygen tetrahedra distortion, rendering it difficult to detect in experiments.

\section{Acknowledgments}
The research was carried out within the state assignment to ISSCM SB RAS (project No. 121032500059-4).
I.S.N. was supported by Russian Science Foundation (grant number 22-73-10206, https://rscf.ru/project/22-73-10206/).
Authors would like to thank Prof. Dr. A.G. Kvashnin for useful discussion.

\section{Data Availability}
Data are available upon reasonable request to the corresponding author.

 \bibliographystyle{elsarticle-num} 
 \bibliography{cas-refs}






\end{document}